# Acoustic Filter Design Using Temperature Tuning

H. SADEGHI, A. SRIVASTAVA, A. V. AMIRKHIZI and S. NEMAT-NASSER

## ABSTRACT

The material properties selection for designing acoustic filters with optimal performance over a range of frequencies requires considerable effort to fabricate and test laboratory samples. To simplify this procedure, one may test a single sample at various temperatures to design an acoustic filter for a desired band-width. The essential idea is to fabricate a single layered periodic elastic composite with constituent materials that have temperature-dependent properties. As temperature is changed, such a composite exhibits a band structure that changes with the change in temperature. Once a desired band structure is attained and the corresponding constituent properties are identified, then new constituents that have those properties at the required temperature can be selected and new sample fabricated. We fabricated a 2-phase composite with periodic layers of polyurea and steel. The temperature is changed from -20°C to 60°C and ultrasonic measurements are performed on the sample over 0.15 to 2.2MHz at each temperature. The first three pass bands are captured experimentally and significant change in the band structure is observed over the test temperature range. Experimental transmission spectrum at each temperature is compared with the theoretical band structure and it is shown that good agreement exists for the observed variation in the band structure.

## INTRODUCTION

In recent years, periodic composites (PCs) have attained considerable attention because of their interesting elastodynamic behavior [1, 2, 3, 4]. These composites display pass bands and stop bands that can be used to design acoustic filters, noise

Hossein Sadeghi, Karagozian and Case, 700 N Brand Blvd., Suite 700, Glendale, CA 91203, USA.
Ankit Srivastava, Department of Mechanical, Materials, and Aerospace Engineering Illinois Institute of Technology, Chicago, IL, 60616 USA.
Alireza V. Amirkhizi, Department of Mechanical Engineering, University of Massachusetts, Lowell, Perry 332, 1 University Ave., Lowell, MA 01854, USA.
Sia Nemat-Nasser, Department of Mechanical and Aerospace Engineering, University of California, San Diego, Mail code 0416, 9500 Gilman Dr., La Jolla, CA 92093, USA.

insulation, and vibration-less environment to house sensitive instruments. Wang et al. [5] studied acoustic wave propagation in a 1-D PC made of layers of copper and epoxy. They used Bloch theory to calculate the theoretical band structure of the composite and showed that their calculation agree with experimental results. Mizuno and Tamura [6] derived an analytical expression for transmission of elastic waves through finite 1-D PCs. They observed that over the stop bands the transmission is very small while over the pass bands a considerable portion of the wave is transmitted. Esquivel-Sirvent and Cocoletzi [7] used transfer matrix method to calculate the band structure of 1-D PCs for both longitudinal and transverse waves. They showed that the reflection coefficient of longitudinal waves measured experimentally in a binary system matches well with transfer matrix calculations. Hussein et al [8] studied the wave propagation in finite and infinite 1-D PCs and showed that with only two unit cells the band structure of the composite can be captured. Amirkhizi and Nemat-Nasser [9] used a general variational method that has been developed by Nemat-Nasser [10] and Nemat-Nasser et al [11, 12] in 1970's for band structure calculation of 1-, 2-, and 3-D elastic composites, to study a 1-D PC made of layers of steel and PMMA. They conducted experiments to verify their results and observed good agreement between theoretical and experimental band structure.

Because of importance of stop bands in PCs, there have been efforts to develop new methods to tune their band structure. Ruzzene and Baz [13] used shape memory alloys to make a 1-D PC and showed that by changing the thermal activation of the shape memory alloy, the width and location of stop bands can be controlled. Goffaux and Vigneron [14] showed that stop bands of a 2-D PC made of square rods distributed in air can be tuned by changing the geometry of the system. They observed that by increasing the rotation angle of the rods, stop bands of the PC can be widened. Zou et al [15] showed that stop bands of a 2-D PC made of rectangular piezoelectric ceramics placed in epoxy matrix can be controlled with changing the filling fraction of the rods. Wu et al [16] used dielectric elastomeric layers in a 1-D PC made of layers of Al and PMMA to make a tunable band-pass acoustic filter. They showed that band structure of the composite can be controlled by changing the applied electric voltage on dielectric layers. Zou et al [17] showed that in-plane stop bands in a 2-D PC made of piezoelectric rods embedded in a matrix can be tuned by changing the out-of-plane wave number. They observed that the width and starting frequency of stop bands can be controlled by changing the filling fraction and the composite's piezoelectricity.

In this paper we show that the band structure of a layered PC can be controlled by changing the ambient temperature, and that this can be used to experimentally find the material properties require for designing an acoustic filter with a required band structure at a desired temperature. For this purpose, a layered PC consisting of layers of steel and polyurea is designed and fabricated. Because polyurea has temperature-dependent mechanical properties, as temperature is changed, the band structure of the composite is expected to change. The change in the stiffness of polyurea with temperature is studied experimentally and used to calculate sample's band structure. Transmission spectrum of the sample is measured using an ultrasonic setup and the changes in the location of the first three pass bands are studied. Theoretical band structure of the sample is compared with experimental data at each temperature to further verify the applicability of this method. Using this method, tunable acoustic filters can be studied for various frequency bands by simply changing the experimental temperature instead of having to use a different sample for each frequency band.

## BAND STRUCTURE OF LAYERED PERIODIC COMPOSITES

Consider longitudinal wave propagation in an infinite 1-D periodic composite. The equation of motion and constitutive relation in 1-D can be expressed as

$$\rho(x)\ddot{u}(x,t) = \frac{\partial}{\partial x}\sigma(x,t), \qquad \sigma(x,t) = E(x)\varepsilon(x,t) \tag{1}$$

where $\rho,\ E,\ \sigma, u$ and $\varepsilon = \partial u/\partial x$ are density, Young's modulus, stress, displacement, and strain, respectively. The exact dispersion relation for longitudinal wave propagation in a 2-phase 1-D PC is given by Rytov [18] as

$$\cos(q(d_1+d_2)) = \cos\left(\frac{\omega d_1}{c_1}\right)\cos\left(\frac{\omega d_2}{c_2}\right) - \Gamma\sin\left(\frac{\omega d_1}{c_1}\right)\sin\left(\frac{\omega d_2}{c_2}\right)$$
$$\Gamma = (1+\kappa^2)/2\kappa, \quad \kappa = \rho_1 c_1/\rho_2 c_2 \tag{2}$$

where $\omega$, $q$, $d_i$, $\rho_i$ and $c_i$ are angular frequency, Bloch wave number, thickness of the *i*-th layer, density of the *i*-th layer, and longitudinal wave velocity of the *i*-th layer (*i*=1,2) in a unit cell, respectively.

## EXPERIMENT

### Sample

Figure 1 shows a unit cell of the sample designed for this study which consists of periodic layers of steel and polyurea (PU). The layers are disks with diameter of 1 in and their thicknesses are given by $t_{st} = 1.15\ mm$ and $t_{PU} = 1\ mm$ for steel and PU, respectively. The longitudinal wave velocity and density of steel are given by $c_{st} = 5130\ m/s$ and $\rho_{st} = 7820\ kg/m^3$, respectively. Density of polyurea is given by $\rho_{PU} = 1100\ kg/m^3$, whereas its longitudinal wave velocity changes with temperature (which is given below). Polyurea samples made for this study are fabricated in UCSD/CEAM lab as described in [19].

### Ultrasonic Measurement

Figure 2(a, b) show a schematic illustration and a photograph of the ultrasonic setup used in this study. A wave packet envelop made of 10 sine waves at the carrier frequency, *f*, multiplied by a half-period of another sinusoidal wave of *1/20 f* carrier frequency is generated through the wave generator and is sent to an amplifier (Figure 3(c)) [9],

$$u(0,t) = Asin(2\pi ft)sin(2\pi ft/20) \quad \text{where } 0 < t < 10/f \tag{3}$$

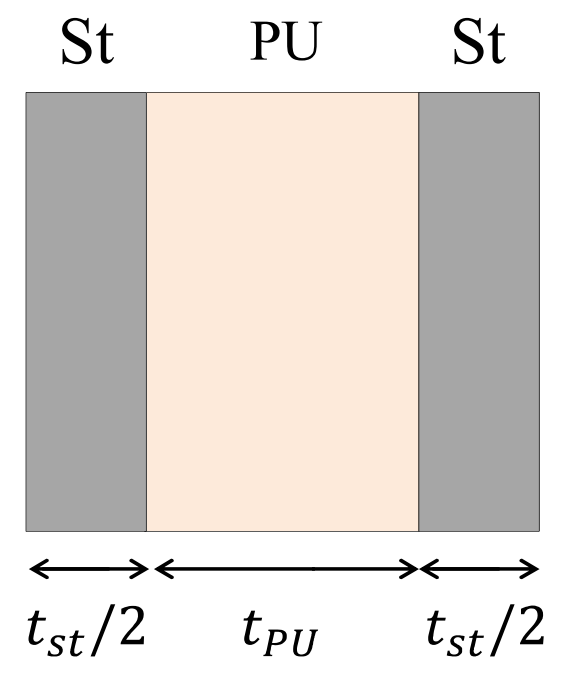

Figure 1: A unit cell of steel/PU layered periodic sample.

The amplified signal is sent to a contact piezoelectric transducer, propagated through the sample, and received by another transducer on the opposite end of the sample. The received transmitted signal is stored by an oscilloscope. The transducers and the sample are kept in a temperature control chamber to maintain the temperature constant during the measurement. The transducers used here are CHRFO18 NDT systems transducers with element diameter of 1 in and nominal center frequency 1.0 MHz. Ultrasonic couplant is used between the transducers and the sample for better transmission of the signal at the interfaces.

In order to measure the viscoelastic properties of polyurea two tests were performed on two samples with different thicknesses, $d_1$ and $d_2$, at each temperature (see Figure 2 (d)). Let the arrival time of the signals transmitted through each sample be $t_1$ and $t_2$ with their amplitudes being $A_1$ and $A_2$, respectively. The longitudinal wave velocity, $c_L$, and attenuation coefficient per unit thickness, $\alpha$, in the material are then given by

$$c_L = \frac{d_2 - d_1}{t_2 - t_1}, \qquad \alpha = \frac{1}{d_2 - d_1} \ln\left(\frac{A_1}{A_2}\right). \tag{4}$$

For a homogenous linear viscoelastic material the real and imaginary parts of the complex modulus can be expressed as [19]

$$E' = \frac{\rho c_L^2 (1 - r^2)}{(1 + r^2)^2}, \qquad E'' = \frac{2\rho c_L^2 r}{(1 + r^2)^2}, \tag{5}$$

where the dimensionless parameter $r$ is given by $r = \alpha c_L / \omega$.

## TEST PROCEDURE AND RESULTS

Figure 3 shows the longitudinal wave velocity and attenuation coefficient per unit thickness for polyurea as a function of temperature at 1.0 MHz. The frequency dependency of material properties of polyurea is studied by Qiao et al. [19]. They showed that in the frequency range that we consider here the change in the properties of polyurea is negligible. Using the temperature dependent material properties of polyurea the theoretical dispersion curve of the steel/PU sample is calculated. In order to study the changes in the band structure, sample is tested in a temperature control chamber and the temperature is varied from -20 ºC to 60 ºC with a step size of 20 ºC. At each temperature the system is left for 15 minutes to reach the state of thermal equilibrium. Ultrasonic measurements are performed over the frequency range from 0.15 to 2.2 MHz and the transmitted signal is recorded at each temperature.

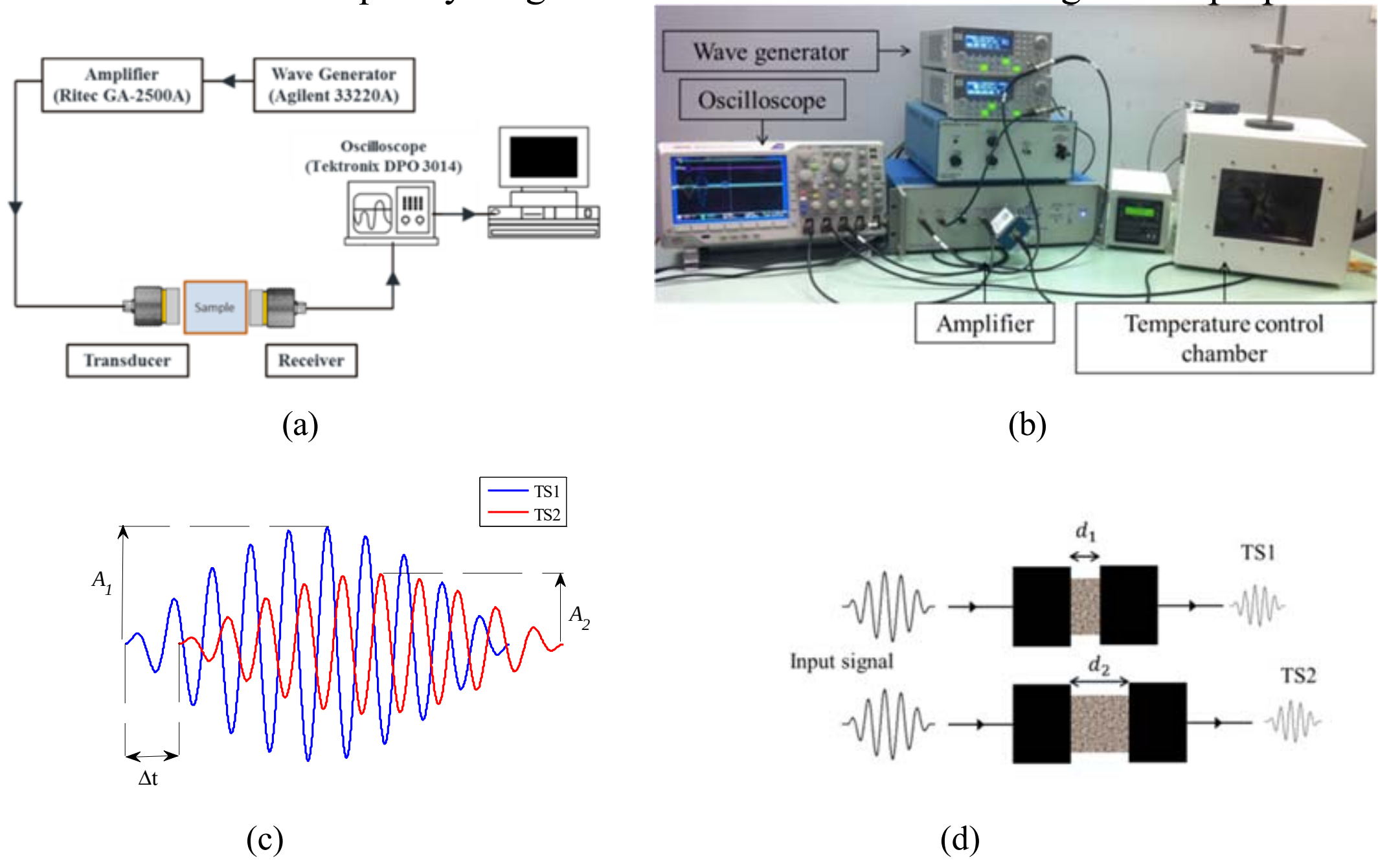

**Figure 2:** (a) Schematic diagram of the ultrasonic setup; (b) ultrasonic setup picture; (c) transmitted signals through the samples with different thicknesses; and (d) sample configurations in ultrasonic tests.

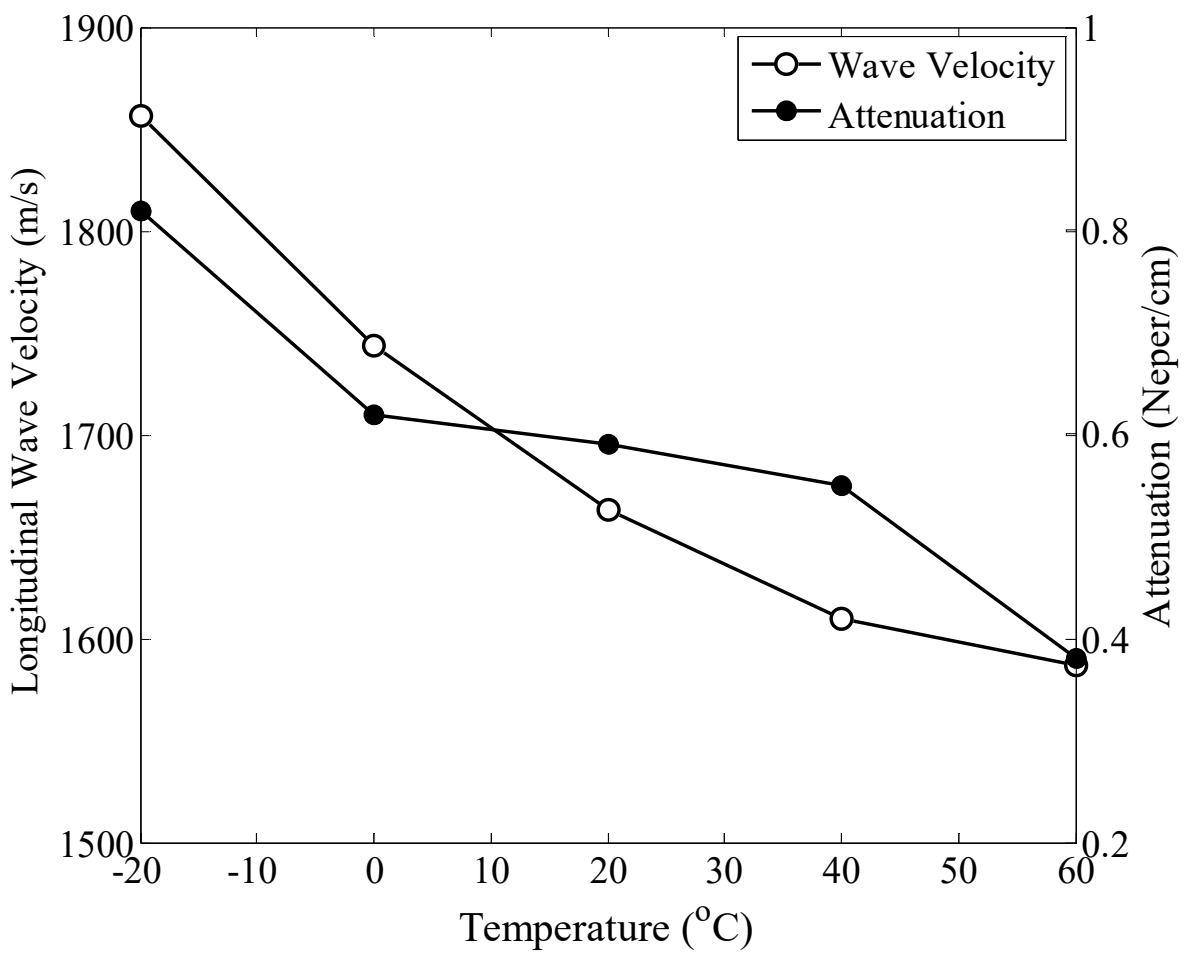

.

Figure 3: Longitudinal wave velocity, and attenuation coefficient per unit thickness of polyurea, as functions of temperature at 1.0 MHz.

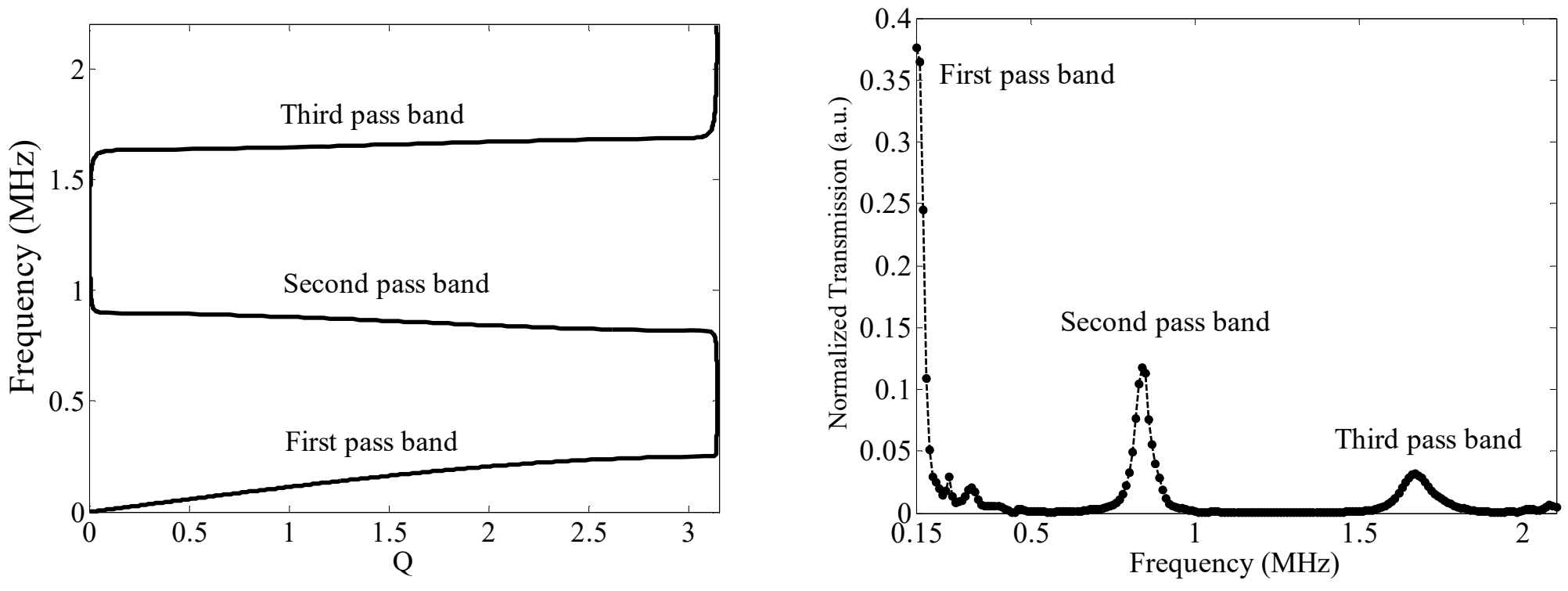

(a) (b)

Figure 4: (a) Dispersion curve of steel/PU sample at T = 20 ºC; and (b) normalized amplitude of transmitted wave through 2 unit cells of the sample at T = 20 ºC.

Figure 4 (a) shows the theoretical dispersion curve of the steel/PU sample at T = 20 ºC where $Q = qd$ is the normalized Bloch wave number and $d$ is the total thickness of the unit cell.

Figure 4 (b) shows the experimental results for normalized amplitude of transmitted wave through 2 unit cells of the sample at T = 20 ºC. The amplitude of transmitted wave is normalized with respect to amplitude of the transmitted wave in a transducer-to-transducer test. It can be seen that there are ranges of frequencies in which the amplitude of the transmitted wave is very small. These are the stop bands of the composite where most of the incident wave is reflected back. In addition, there are ranges of frequencies in which a considerable portion of the incident wave is transmitted through the sample. Those are pass bands of the composite. In the frequency range that measurements are performed, the first three pass bands of the composite are captured. Furthermore, it is observed that the value of the transmitted signal at higher pass bands is smaller compared to the lower pass bands, which is due to higher viscous dissipation at higher frequencies.

Figure 5 shows the normalized amplitude of the transmitted wave as a function of frequency through 2 unit cells of the steel/PU sample at different temperatures. It can be observed that as the temperature is changed from -20 ºC to 60 ºC the band structure changes significantly. Specifically, the pass bands move to lower frequencies when the temperature is increased. This is attributed to the reduction in the elastic stiffness of polyurea as the temperature is increased.

Figure 6 shows the comparison between the theoretical and experimental results for frequencies at the beginning of the 1st pass band ($f_1^{ex}$, $f_1^{th}$) as well as at the center of the 2nd and 3rd pass bands ($f_2^{ex}, f_2^{th}, f_3^{ex}, f_3^{th}$). It can be seen that experimental results match well with theoretical calculations.

## CONCLUSION

It is shown that the band structure of layered PCs can be controlled by changing the ambient temperature. Ultrasonic measurements are performed at different temperatures on a sample made of periodic layers of steel and polyurea. It is shown that the band structure of the sample changes significantly as the temperature is changed. Using this, tunable acoustic filters can be studied for various frequency bands by simply changing the experimental temperature instead of having to use a different sample for each frequency band. Temperature dependent elastic properties of polyurea are measured experimentally and used to calculate the band structure of the sample. It is observed that the calculated band structures are in good agreement with the experimental results at different temperatures. This shows acoustic filters can be designed at a target frequency with a desired band-width through reverse engineering once the corresponding properties are identified using here proposed experimental tool.

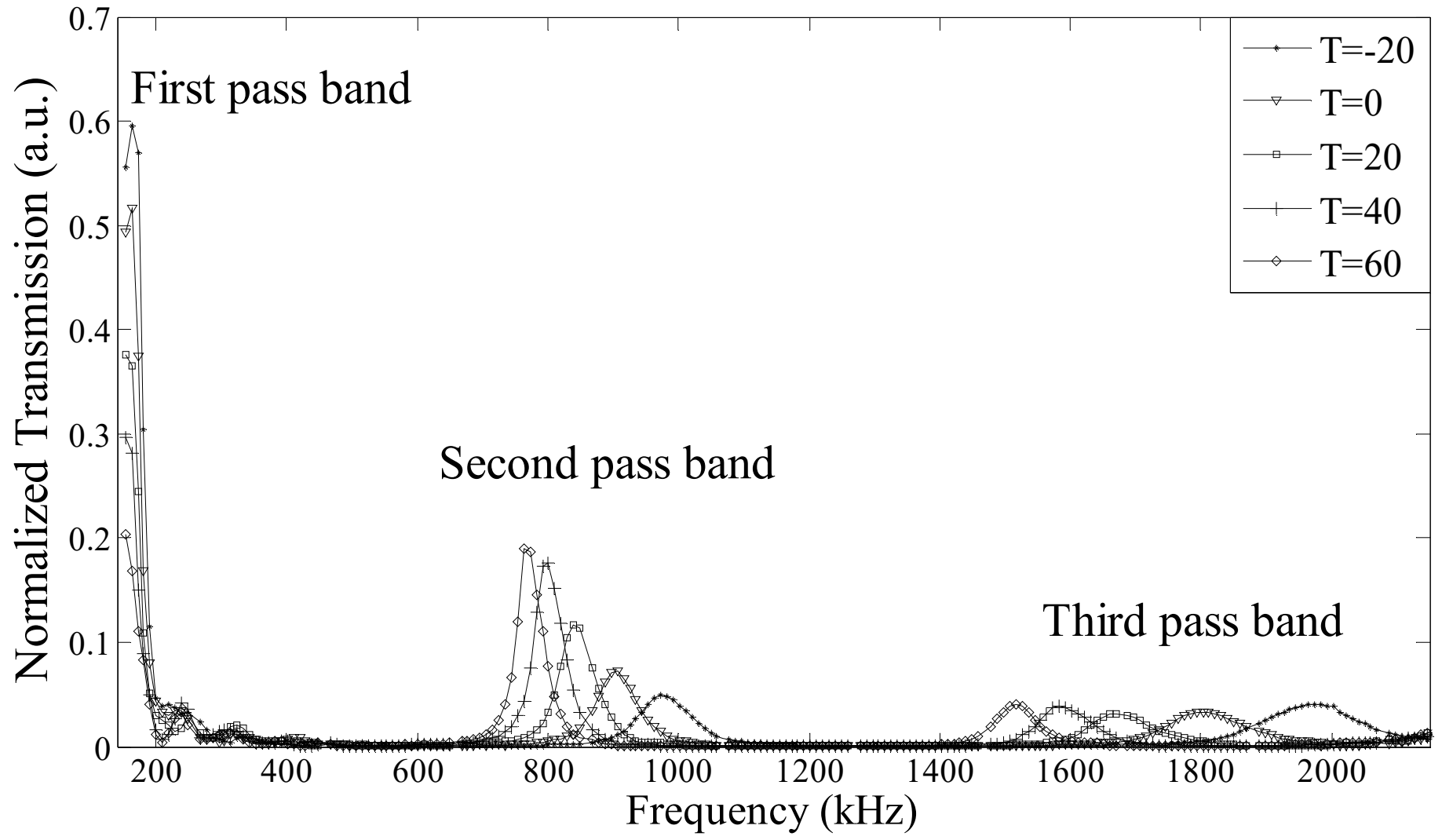

Figure 5: Normalized amplitude of transmitted wave through 2 unit cells of the sample as a function of frequency at different temperatures.

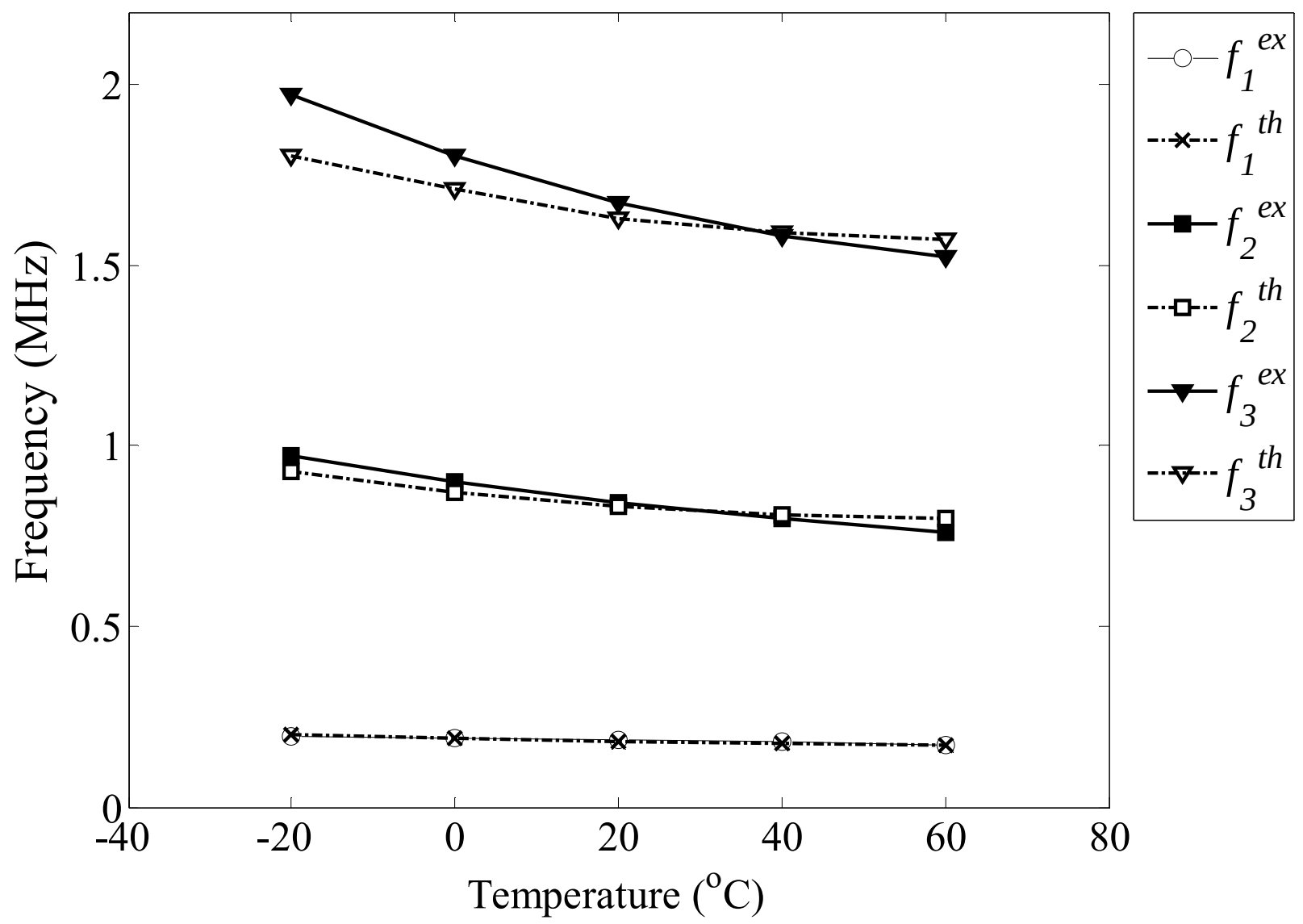

Figure 6: Comparison of the experimental and theoretical frequency at the end of the 1st pass band as well as at the center of the 2nd and 3rd pass bands of the PU/steel sample at different temperatures.

## ACKNOWLEDGEMENT

This research has been conducted at the Center of Excellence for Advanced Materials (CEAM) at the University of California, San Diego, under DARPA Grant RDECOM W91CRB-10-1-0006 to the University of California, San Diego.